\documentstyle[aps]{revtex}
\begin{document}
\draft
\title{Two-component radiation model
of the sonoluminescing bubble}
\author{David Tsiklauri}
\address{Physics Department, University
of Cape Town, Rondebosch 7700, South Africa and
Department of Theoretical
Astrophysics, Abastumani Astrophysical Observatory,
2$^A$ A. Kazbegi ave., Tbilisi 380060, Republic of Georgia.
Internet: tsiklauri@physci.uct.ac.za}
\date{\today}
\maketitle
\begin{abstract}
Based on the experimental data from Weninger, Putterman \& Barber,
Phys. Rev. (E), {\bf 54}, R2205 (1996), we offer an alternative
interpretation of their experimetal results.
A model of  sonoluminescing bubble which
proposes that the electromagnetic radiation originates from
two sources: the isotropic black body or bramsstrahlung emitting  core
and
dipole radiation-emitting shell of accelerated electrons driven by the
liquid-bubble interface is outlined.
\end{abstract}


Theoretical explanation of the sonoluminescence has been a long
standing puzzle since 1934 when it was observed for the first
time [1]. The most viable theoretical models of the phenomenon
are based on so called shock wave model [2] which is capable to
explain certain characteristic features of the effect.
However, this model is constrained by the assumption of the
spherical symmetry of the bubble during all stages of its
collapse.

However, recent pioneering experimental studies [3] convincingly
showed
existence of an emission component with dipole angular distribution of
intensity which strongly suggests presence of some kind of
non-spherical
dynamics of the bubble.

Angular dependence in the intensity of sonoluminescene can be
discribed by the following correlation [3]
$$
\Delta Q_{AB}(\theta_{AB})={1 \over{ \bar{Q}_A \bar{Q}_B }}
\biggl< \left({Q_A(i)- \bar{Q}_A }\right)
\left({Q_B(i)- \bar{Q}_B }\right) \biggr>_i, \eqno(1)
$$
where $\theta_{AB}$ is the angle formed by the photomultiplier
tubes $A$ and $B$ and the bubble which is positioned at the
vertex. $Q_A(i)$ is the total charge recorded in the
detector $A$ on the $i$-th flash, $\bar{Q}_A$ is the running
average of $Q_A(i)$ and $<>_i$ denotes an average over $i$.
Major experimental results obtained by the authors of Ref.[3]
are as follows:

(i) detection of two light emission components with isotropic
and dipole angular distribution through mesurement of
$\Delta Q_{AB}(\theta_{AB})$.

(ii) Finding of qualitatively different physical states of
the sonoluminescing bubble in which the two emission components
have different share in total light intensity.

(iii) measurement of intensity fluctuations in the different
physical states of sonoluminescing bubble.

(iv) measurement of the correlation $\Delta Q_{AB}$ given
by Eq.(1) as a function of time delay $\Delta t$ between
acquisitions in detectors $A$ and $B$.

Authors of Ref.[3] interpreted their  experimental
results (presence of the dipole emission component) as
due to the refraction of light by the non-spherical
bubble wall i.e. liquid-bubble interface.
Thier major argument was that red light ($\lambda > 500$ nm)
showed no angular correlation. Whereas, blue light
(260 nm $< \lambda < $  380 nm) was significantly correlated.
This exprimental fact was interpreted [3] as dominance of
diffraction over refraction in the case of long wavelength
(since the radius of bubble is about the same size as red light
wavelength) and vice versa in the case of short wavelength (blue
light).

Below we show that these novel experimental facts can be explained
in an alternative way and outline fundamentals of the
two component model.

Explanation of presence of the dipole component in terms
of the light refraction from the non-spherical liquid-bubble
interface [3] implies the primary isotropic core emission
comes from a point source which is more likely to be the
either black body radiation coming from the contents of the
bubble which was heated up by the implosion [4] or
bremsstrahlung emitted from the air after it has been ionzied
by shock compression [2].
Further, light from this point source is refracted from
the non-spherical liquid-bubble interface which results in
a dipole angular distribution of the detected light [3].
However, it is reasonable
to assume that the angularly correlated component primarily has a
dipole origin itself. Preliminary numerical simulations showed that
liquid-bubble interface achieves substantial accelerations
at the final stages of the collapse. The measure of the
latter physical quantity could be $\ddot{R}(t)$
(second derivative of the radius with respect to time)
calculated from the Reyleigh-Plesset equation
which even in $adiabatic$ calculation acquires values
$\sim 10^{16}$m$/$sec$^2$.
Similar result yields more rough estimate:
$a \sim \Delta v/ \Delta t \sim 2 v / \Delta t$,
where $v$ is the maximal velocity acquired during the collapse
($\sim 5$ km/sec) and $\Delta t$ is the time-scale of the radius
turnaround ($\sim$ psec).
The free electrons which
come from ionization of the air will be easily  dragged
by the liquid-bubble interface since they have small inertia.
One could safely assume that typical accelerations of the free
electrons dragged by the liquid-bubble interface will be order
of the $\ddot{R}(t)$. It is well known that  accelerated, charged
particles moving with non-relativistic velocities (which is
apparently
the case for particles within the sonoluminescing bubble)
emitt dipole radiation. However, a {\it spherical} shell of electrons
driven by liquid-bubble interface will not emit dipole radiation
since the dipole moment of such configuration is zero
(unless the electrons are non-uniformly distributed
on the interface, which is quite improbable).
Previous experimental studies (involving light scattering techniques
along with relevant Mie-scattering algorithms) suggest that
the bubble remains spherically symmetric until the final
stages of the collapse and only then (presumably on the psec
time-scale)
it becomes distorted by "shape" instabilities [4].
Therefore, at this very instance of time, dipole moment of
the system suddenly becomes non-zero.
Thus, allowing dipole radiation to take place.

It is important to note that the two component model
in which dipole component originates from the dipole
emission of shell of electrons dragged by liquid-bubble interface
is consistent with the  experimental fact [3] that the red light has
no angular correlation whereas blue light shows significant angular
correlation. It is known that the spectral resolution
of the intensity of dipole radiation is given by [5]
$$
{d E_\omega}={{{4 \omega^4}\over{3 c^3}}|{\bf d}_\omega|^2
{{d \omega}\over{2 \pi}} \propto \omega^4      }.  \eqno(2)
$$
Therefore, since the intensity of the dipole radiation
strongly depends on frequency (via Eq.(2)) for low frequencies
(red light) intensity of dipole radiation is overwhelmed
by the isotropic core (black body or bremsstrahlung) emission,
whareas in the case of high
frequencies (blue light) dipole radiation is more pronounced.

As we mentioned above yet another significant experimental result
of Ref.[3] is the
measurement of the angle dependent correlation $\Delta Q_{AB}$,
(see Ref.[3] for details) as a function of a time
delay $\Delta t$ between acquisitions in photomultiplier tubes
$A$ and $B$. This data is important because it provides a clue
to determine a source of dipole component.
In particular, it has been shown [3] that angle dependent
correlation  $\Delta Q_{AB}(\Delta t)$ reveals a long time delay
which indicates that dipole component is due to the peculiarities
of hydrodynamic motion.
After excluding various possibilities
authors of Ref.[3] concluded that the most viable mechanism
is refraction of light by {\it non-spherical} liquid-bubble
interface. Therefore, non-sphericity of the bubble plays a key
role in their scenario. However, this argument would also perfectly
fit our alternative interpertation of the experimental data,
because this is the non-sphericity of the bubble
which makes the dipole moment of the shell of electrons driven
by the liquid-bubble interface non-zero. Thus allowing the system
to emit dipole radiation.

It is also important to address issue
of the intensity fluctuations. In ref.[3] it was established that
sonolumenscent states where dipole component dominates isotropic
component exhibit very large fluctuations in emission intensity.
Sonoluminescing states with
fraction of dipole component six parts per thousand peak
to peak are characterized by intensity fluctuations that are over a
factor of 10 greater than states with dipole components of about
one part per thousand or less. clue to the explanation of this effect
could lay in non-sphericity of the bubble at instance of light
emission.
Sonoluminescing state with high fraction of dipole component is
achieved
when there are large deviations from spherical shape of the bubble.
Because of this process is caotic (since it draws its orgin from
some kind
of hydrodymic instability) and the position of the photomultipiler
tube is fixed this results in large fluctuations of intensity.
This explanation is equally valid for the refraction model [3] and
our two component model, since in both of them cause of dipole
emission ultimately is non-sphericity of the bubble.

While mentioning isotropic core emission above, we referred
to the black body and bramsstrahlung radiation in an equal manner.
However, as we shall see below, thanks to the discovery
of the two different sonoluminescing states [3] with no (small)
and dominant
dipole components, futher experimetal measurements of the
sonoluminescing flash duration could discriminate between
black body and bremsstrahlung emission mechanisms as well
as between the refraction model [3] and our two component model.
As it was emphasized in ref.[4] black body radiation model
predicts that the duration of the sonoluminescence light flash
should be order of tens of nsec because the temperature
of the contents of the bubble is order of 2000K
and larger for a time span over 20 nsec.
On the other hand, detailed numerical simulations of the shock
wave model based on the bremsstrahlung emission assumption
confirms tens of psec duration flash [2].
The refraction model [3] which explains presence of dipole
component in certain sonoluminescing states apparently
will never predict change in the duration of the light flash,
since the light is simply refracted from non-spherical liquid-bubble
interface. Whereas, in our two component model this is possible
because dipole component has different origin --- dipole radiation of
the accelerated electrons driven by the liquid-bubble interface.
Sonoluminescing state in which dipole component is dominant
and the core black body emission is assumed deserves
particular attention, because in this case our two component model
predicts different light flash duration.
Dominance of the dipole component in our model means that
the core isotropic component (black body radiation) has very
low intensity and all detected light comes from the dipole radiation
of shell of eletrons driven by the liquid-bubble interface.
As we mentioned above, dipole emission of such configuration is
possible when the bubble looses spherical shape (when the dipole
moment suddenly becomes nonzero) which happens
at the very final stages of the collapse, presumably on the
psec timescale.
Apparently, in this case light refraction model [3] would still
predict tens of nsec duration flash
since the primary (and the only) emission source
is the isotropic black body radiation
and of course refraction cannot change the duration of the light
flash itself. 
Authors of Ref. [3]
established that the both states (with no (small) and dominant
dipole components)
exhibit the same flash to flash synchronicity. However, they have
not presented measurements for duration of the light flash in
both cases.
This is important, beacuse under assumption of black body core
emission it would allow to discriminate between
refraction model and our model.
In the case of sonolminescing state with no (small) dipole component
both models would predict the same duration of the light flash
which would be order of tens of nsec, since in both cases emission
comes from isotropic black body source which has relatievly
large time scale [4].
Whereas, in the case of sonoluminescing state with dominant dipole
component our model would predict short (tens of psec) light flashes
and refraction model would still predict long (tens on nsec) flashes.
On the other hand assuming that isotropic core emission is of
bremsstrahlung
type both the refraction [3] and two component model predict the
same flash durations tens of psec.
We ought to remark that in the literature duration of light flash
is claimed
to be tens of psec. To the best knowledege of the author, source
reference for this information is Ref. [6]. However, awareness
of existence of the
dipole component emerged from later experimental studies
presented in Ref. [3].
Therefore, {\it a priori} it is unclear whether measured duration
of the flash
tens of psec [6] was for the state with dominant dipole component or
 with small
one. To clarify this point further expremintal studies are necessary.

Finally, we conlude with estimate of the peak power of the
sonoluminescence radiation based on the assumption that the all light
comes from dipole radiation of shell of electrons dragged
by the liquid-bubble interface (dominant dipole state).
Typical value of the peak power
is order of tens of mW [2]. We know that the total power
of dipole radiation of system of accelerated electrons
emitted in every direction is [5]
$$
I={{2}\over{3 c^3}} {\bf \ddot{d}}^2,\eqno(3)
$$
where ${\bf \ddot{d}} \equiv \sum e {\bf \ddot{r}}$ denotes the
second derivative of the total dipole moment
with respect to time ($\bf r$ stands for a radius vector
of a particular electron). Apparently, it is impossible
to estimate $I$ unless angular distribution and the acceleration
of every electron on the non-spherical shell is know.
However, assuming that we have system of $N$ electrons
moving with plausible acceleration value
$a \sim 10^{16}$m$/$sec$^2$ in the same direction
(Here we mention that there are models which propose
formation of a jet at the final stages of the collapse (see further
Refs. [13-14] in Ref.[3])) and putting $\sim 10^{26}$ for the $N$
we end up with reasonable value for the peak power.

As it was argued above, present experimental data allows
alternative interpretation.
Therefore it is important to perform new experimental
measurements of the light flash duration in the two sonoluminescing
states
with dominant and no (small) dipole components in order to
discriminate between the refraction model [3] and our two component
model as well as between black body and bremsstrahlung
emission mechanisms. Table 1, where we present
anticipated durations of the sonoluminescing flash
by the refraction [3] and our two component models under the
assumption of the black body and bremsstrahlung core emission
mechanisms
summarizes specific predictions of the models.

\newpage
Table 1. predicted durations of the sonoluminescing flash
by the refraction and two component models under the
assumption of the black body and bremsstrahlung core emission
mechanisms respectively.
\newpage
\centerline{\bf Table 1}
\vskip 7cm
\begin{tabular}{|c|c|c|c|} \hline
{\bf Core emission} & {\bf SL  state} & {\bf Dipole emission model}
& {\bf Flash duration}  \\ \hline
 & no (small) dipole & refraction & tens of nsec \\ \cline{3-4}
 black body & & two component & tens of nsec \\ \cline{2-4}
  & dominant dipole & refraction & tens of nsec \\ \cline{3-4}
  & & two component& tens of psec \\ \cline{1-4}
 & no (small) dipole & refraction & tens of psec \\ \cline{3-4}
 bremsstrahlung & & two component & tens of psec \\ \cline{2-4}
  & dominant dipole & refraction & tens of psec \\ \cline{3-4}
  & & two component & tens of psec \\ \cline{1-4}
\end{tabular}

\end{document}